\documentclass[doublecol]{epl2}
\usepackage{graphicx}
\usepackage{nicefrac}

\title{Phase transition in the Sznajd model with independence.}

\author{K. Sznajd-Weron\inst{1}, M. Tabiszewski\inst{1}, A. M. Timpanaro\inst{2}}
\shortauthor{K. Sznajd-Weron \etal}

\institute{                    
  \inst{1} Institute of Theoretical Physics, University of Wroc{\l}aw\\
  \inst{2} Instituto de F\'isica, Universidade de S\~ao Paulo
}


\abstract{
We propose a model of opinion dynamics which describes two major types of social influence -– conformity and independence. Conformity in our model is described by the so called outflow dynamics (known as Sznajd model). According to sociologists' suggestions, we introduce also a second type of social influence, known in social psychology as independence. Various social experiments have shown that the level of conformity depends on the society. We introduce this level as a parameter of the model and show that there is a continuous phase transition between conformity and independence. 
}

\begin{document}

\maketitle

\section{Introduction}
Opinion dynamics is one of the most studied subjects in the field of sociophysics. Among a number of models that have been proposed (for a recent review see \cite{CFL2009}), simple models based on Ising spin variables are particularly interesting. In all these models, from the voter model \cite{HL1975}, majority rule \cite{G2002,KR2003} to the Sznajd model \cite{SWS2000}, opinions are described by discrete variables $S = \pm 1$. The ferromagnetic state is an attractor for all three models in one and two dimensions, as well as in the case of complete graphs \cite{CFL2009}. Obviously, in real social systems complete unanimity is never reached. Moreover, in real systems the public opinion does not reach any fixed point and permanently changes. To make models of opinion dynamics more realistic, Galam has proposed two modifications:
\begin{itemize}
\item Contrarian behavior \cite{G2004} -- with a certain probability an agent adopts the choice opposite to the prevailing choice of the others, whatever this choice is.
\item Inflexibles \cite{GJ2007} -- inflexible agents keep their opinion always unchanged.
\end{itemize}  
As shown by Galam, for a low concentration of contrarians a new mixed phase is stabilized, with a coexistence of
both opinions, i.e. minority persists. Moreover, there is a phase transition into a new disordered phase with no dominating opinion. In the case of the Sznajd model, contrarian behavior has been studied for the first time by de la Lama et al. \cite{LLW2005} and the same results have been obtained. As suggested by Galam \cite{G2004}, these results may be put in parallel with ``hung elections'' in America (2000) and Germany (2002). 


In the field of social psychology, contrarian behavior, introduced by Galam in \cite{G2004}, is nothing more than anti-conformity -- a particular type of nonconformity \cite{Nail_2000}. There are two widely recognized types of nonconformity: anti-conformity and independence. From a social point of view, it is very important to distinguish between independence and anti-conformity \cite{SBAH2006}. The term `independence' implying the failure of attempted group influence. Independent individuals evaluate situations independently of the group norm. On the contrary, anti-conformists are similar to conformers in the sense that both take cognizance of the group norm -- conformers agree with the norm, anticonformers disagree. As noticed in \cite{SBAH2006}: \emph{This behaviour is a bit of a paradox, because in order to be vigilant about not doing what is expected, one must always be aware of what is expected. In contrast, truly independent people are oblivious to what is expected}. Numerous studies have shown that the level of conformity/anti-conformity depends on the society (culture, age, etc.) \cite{SBAH2006,Myers2010}.

In this paper we examine the influence of independence on the Sznajd model in one and two dimensions, as well as on a complete graph. We show that contrarians are not needed in order to obtain phase transitions as in \cite{G2004,LLW2005} and similar results can be observed in the presence of independent behavior. Below a certain critical independence value $p_c$, the minority opinion coexists with the majority and above $p_c$ there is no majority in the society, the system is in the so called stalemate state. 

To generalize our model and make it more realistic from a social point of view. we introduce also a flexibility factor $f$, which describes what is the probability of an opinion change in the case of independent behavior. We show, both analytically (for complete graphs) and using Monte Carlo simulations, that the critical threshold of independence $p_c$ decays with the flexibility factor $f$. This means that in an inflexible (conservative) society the critical independence factor is high. This result implicates that in conservative societies, even if the level of independence is high, there is always a majority in the system -- in the case of democratic voting, one of the two options wins. On the contrary, in highly flexible societies the stalemate situation is highly probable. Our result is one of the possible explanations why in modern societies (in which "tradition" is not as valuable as in old days), status quo situations happen more and more often. 

\section{Model}
We consider a set of $N$ individuals, which are described by the binary variables: $S=1$ ($\uparrow$) or $S=-1$ ($\downarrow$). At each elementary time step, a group of people is chosen randomly and it influences its surrounding individuals. In the original model \cite{SWS2000} only one type of social influence (conformity) was considered: 
\begin{enumerate} 
\item
On a complete graph, two individuals are chosen at random and they influence a third randomly chosen individual \cite{SL03}. 
\item
In one dimension (1D), a pair of neighboring individuals $S_iS_{i+1}$ is chosen and it influences two neighboring sites $S_{i-1}, S_{i+2}$. In this paper, to be consistent with the case of a complete graph, we will use the modified version in which only one of the two (left $S_{i-1}$ or right $S_{i+2}$), chosen randomly, will be changed. This kind of modification has been introduced for the first time by Slanina \cite{SSP2008}. 
\item
Several possibilities of generalization to the square lattice were proposed by Stauffer et al \cite{SSO2000}. Here we use probably the most popular rule - a $2 \times 2$ panel of four neighbors is chosen randomly and influences its surroundings. In this paper we use a modified version to be consistent with the rules above -- only one from the 8 neighbors of the panel is randomly chosen to be changed. 
\end{enumerate}

In real social systems, conformity is one of the most recognized types of social response. There are many factors that affect the likelihood of conformity, among them culture is one of the most important. For example, the level of conformity (and simultaneously anti-conformity) is much higher in Japan than in America \cite{SBAH2006}. The well known American slogan ``Do your own thing'' reflects a tendency to independent behavior. Therefore, in this paper we introduce a second type of social response, known as independence. With probability $p$, an individual $S_k$ chosen to be changed will not follow the group, but act independently -- with probability $f$ it will flip, i.e. $S_k \rightarrow -S_k$ and with probability $1-f$ stay unchanged, i.e $S_k\rightarrow S_k$. The parameter $f$ is called flexibility, since it describes how often an individual will change its opinion in case of independent behavior. With probability $1-p$ the individual will follow the usual Sznajd conformity rules, described above. 

\section{Model on a Complete Graph}
We consider a set of $N$ Ising spins $S_i=\pm 1, \; i=1,\ldots,N$ on a complete graph. In each elementary time step $t$ two spins $S_i$ and $S_j$ are chosen randomly. They will influence a third randomly chosen spin $S_k$ in the following way: 
\begin{itemize}
\item conformity (original Sznajd rule), with probability $1-p$: $S_k(t+dt)=S_i(t)$ if $S_i(t)=S_j(t)$, otherwise $S_k(t+dt)=S_k(t)$,
\item independence, with probability $p$: $S_k(t+dt)=-S_k(t)$, with probability $f$ or $S_k(t+dt)=S_k(t)$, with probability $1-f$.
\end{itemize}
Time $t \rightarrow t+1$ after $N$ elementary time steps, i.e. $dt=1/N$. 

In the case of a complete graph, the state of the system is completely described by the magnetization (or public opinion from a social point of view) defined as:
\begin{equation}
m(t) = \frac{1}{N} \sum_{i=1}^N S_i(t).
\end{equation}

\begin{figure}
\begin{center}
\includegraphics[scale=0.4]{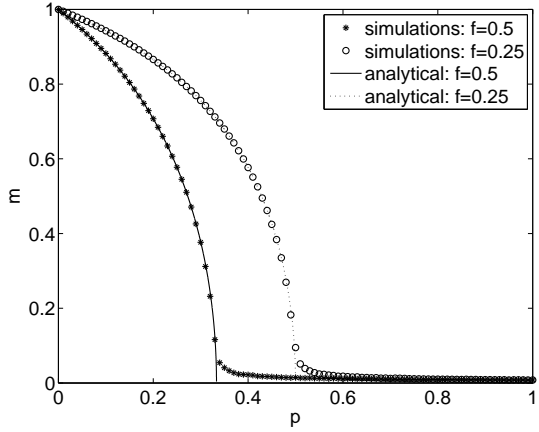}
\caption{Stationary value of magnetization for two values of the flexibility factor, $f=0.5$ (solid line) and $f=0.25$ (dotted line), from analytical calculations and from Monte Carlo simulations for the system size $N=10^3$. Averaging was done over $10^3$ samples.}
\label{analit}
\end{center}
\end{figure}

Let us denote by $N_{\uparrow}$ the number of spins `up' and by $N_{\downarrow}$ the number of spins `down':
\begin{eqnarray}
N_{\uparrow}+N_{\downarrow}& = & N \nonumber\\
N_{\uparrow}-N_{\downarrow}& = & mN.
\label{eq_NpNm}
\end{eqnarray} 
From the equations above we can easily derive the formula for the probability of choosing randomly a spin $S=+1$:
\begin{eqnarray}
P_+(t) & = & \frac{N_{\uparrow}}{N} = \frac{1+m(t)}{2}. 
\label{mag_p}
\end{eqnarray} 
The evolution of the system is described by the following equation:
\begin{equation}
P'_+ - P_+ = pfP_- - pfP_+ + (1-p)P_+^2P_- - (1-p)P_-^2P_+, 
\end{equation}
where we use the notation $P'_+ \equiv P_+(t+dt)$ and $P_+ \equiv P_+(t)$ and the probability of choosing a 'down' spin is $P_-=1-P_+$.
We look for fixed points of the above transformation:
\begin{eqnarray}
P'_+ - P_+ & = &(1-2P_+)\left[(1-p)P_+^2 -(1-p)P_+ + pf\right]=0. 
\end{eqnarray}
The equation above has the following solutions:
\begin{eqnarray}
P_+^0=1/2 \; &\mbox{for} \; p \in [0,1] \nonumber\\
P_+^{1,2}=\frac{1-p \pm \sqrt(\Delta)}{2(1-p)} \; &\mbox{for} \; p <\frac{1}{1+4f},
\end{eqnarray}
where $\Delta=(1-p)(1-p-4pf)$.
Fixed points for the magnetization can be easily calculated from relation (\ref{mag_p}) and are presented in Figure \ref{analit}. There is a continuous phase transition at $p_c=1/(1+4f)$ -- for $p<p_c$ minority coexists with majority and for $p>p_c$ there is a stalemate (status-quo) situation. To confirm our analytical results, we have provided Monte Carlo simulations on a complete graph, for several lattice sizes. Initially the system has been ferromagnetically ordered ($m(0)=1$) and then evolved according to the algorithm described in this section. These results are presented in Figure \ref{analit} and agree with the analytical prediction.

\section{Scaling}
\begin{figure}
\begin{center}
\includegraphics[scale=0.4]{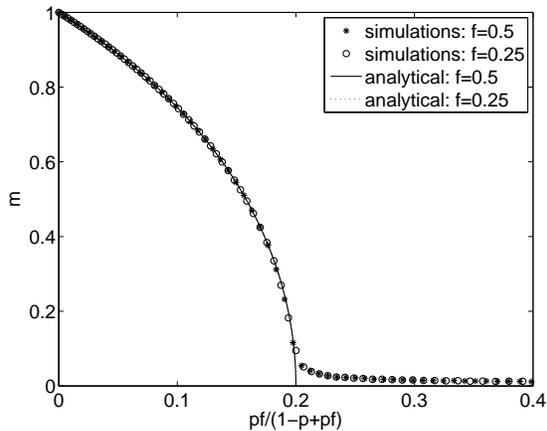}
\caption{Rescaled results from Fig. \ref{analit}.}
\label{analit_scal}
\end{center}
\end{figure}

Although the model considered in this paper has 2 parameters, independence $p$ and flexibility $f$, it can be shown that in fact results depend on the ratio $pf/(1-p+pf)$ (see Figure \ref{analit_scal}). To show this scaling let us first sum up the rules of the model as:
\begin{description}
\item[Rule I] the usual Sznajd rule,
\item[Rule II] a site is fliped ($S_i\rightarrow -S_i$),
\item[Rule III] nothing happens.
\end{description}
At each iteration, we choose a group of sites that will attempt to convince one of their neighbours, chosen at random, then one of the 3 rules above is chosen at random to be followed by this chosen neighbour, with the rules I, II and III being followed with probability $1-p$, $pf$ and $p(1-f)$ respectively. Firstly, we note that rule III doesn't change the state of the system (the opinions of each of the sites) and that the parameters $p$ and $f$ are not important in a given iteration, once the rule to be followed is chosen. Suppose now that we keep records of the state of the system after the iterations where either rule I or rule II was followed. These records would form a sequence of states, whose statistical properties depend only on the initial conditions and on the ratio between the probabilities of following rule I and II:
\begin{equation}
r = \frac{pf}{1-p}.
\end{equation}
So if two models have parameters $(p,f)$ and $(p',f')$ such that $r = r'$, then they would both generate statistically similar sequences, given the same initial conditions. In order to compare both models, we need to find the mean time for a change $\tau$ (measured in iterations) as a function of $p$ and $f$. We follow either rule I or rule II with probability $\tilde{p} = 1-p+pf$, so:
\begin{eqnarray}
\tau & = &  \tilde{p} + 2\tilde{p}(1-\tilde{p}) + 3\tilde{p}(1-\tilde{p})^2 + \ldots \nonumber\\
& = & \tilde{p}((1 + (1-\tilde{p}) + (1-\tilde{p})^2 + \ldots) \nonumber\\
& + & ((1-\tilde{p}) + (1-\tilde{p})^2 + (1-\tilde{p})^3 + \ldots) + \ldots) \nonumber\\
& = &  \tilde{p}\left(\sum_{k=0}^{\infty} (1-\tilde{p})^k \left(\sum_{q=0}^{\infty} (1-\tilde{p})^q \right)\right) = \frac{1}{\tilde{p}}. 
\end{eqnarray}
Therefore:
\begin{equation}
\tau(p,f) = \frac{1}{1-p+pf}.
\end{equation}
To compare the models with parameters $(p,f)$ and $(p',f')$ we must look at time instants $t$ and $t'$, such that
\begin{equation}
\frac{t}{\tau(p,f)} = \frac{t'}{\tau(p',f')}.
\end{equation}
Finally, we are interested in the behavior of the magnetization $m(t,p,f)$ of the model with parameters $(p,f)$ at the time instant $t$. From
\begin{eqnarray}
\left\{
\begin{array}{ccc}
t(1-p+pf) & = & t'(1-p'+p'f') \\
\nicefrac{pf}{(1-p)} & = & \nicefrac{p'f'}{(1-p')}
\end{array}
\right. 
\end{eqnarray}
follows:
\begin{eqnarray}
m(t,p,f) = m(t',p',f').
\end{eqnarray}
If we pick $f'=1$, then
\begin{eqnarray}
\left\{
\begin{array}{l}
t' = t(1-p+pf) \\
p' = \nicefrac{pf}{(1-p+pf)}
\label{scal}
\end{array}
\right..
\end{eqnarray}
So defining $M(t,p) \equiv m(t,p,1)$ it follows that
\begin{eqnarray}
m(t,p,f) = M\left(t(1-p+pf),\frac{pf}{1-p+pf}\right),
\end{eqnarray}
and for each value of $t$, plotting 
\begin{equation}
m(\nicefrac{t}{(1-p+pf)}, p, f)\times\nicefrac{pf}{(1-p+pf)}
\end{equation} 
should collapse the curves, as
\begin{eqnarray}
m\left(\frac{t}{1-p+pf},p,f\right) = M\left(t,\frac{pf}{1-p+pf}\right).
\end{eqnarray}
We have used the scaling given by Eq. (\ref{scal}) and indeed the data collapses in the case of a complete graph (see Fig.\ref{analit_scal}). In the next sections we will see that the scaling derived here is valid also in the case of one- and two-dimensional lattices.

In figure \ref{analit_scal} we can see a phase transition between the situation where a majority exists and the stalemate situation. This transition happens for a value of independence $p_{c}$, that depends on the flexibility value $f$. We can apply this scaling to find that this dependence must be of the form

\begin{equation}
p_{c} = \frac{1}{1+\alpha f},
\end{equation}

where, if $P_{c}$ is the critical independence for $f=1$, then $\alpha$ is given by

\begin{equation}
\alpha = \frac{1}{P_{c}} - 1.
\end{equation}

This means that if there is a phase transition, then the critical value of independence decreases with increasing flexibility. We will give some interpretations for this further on, when the model in a two-dimensional lattice is analyzed.

\section{Model on a one-dimensional lattice}
\begin{figure}
\begin{center}
\includegraphics[scale=0.4]{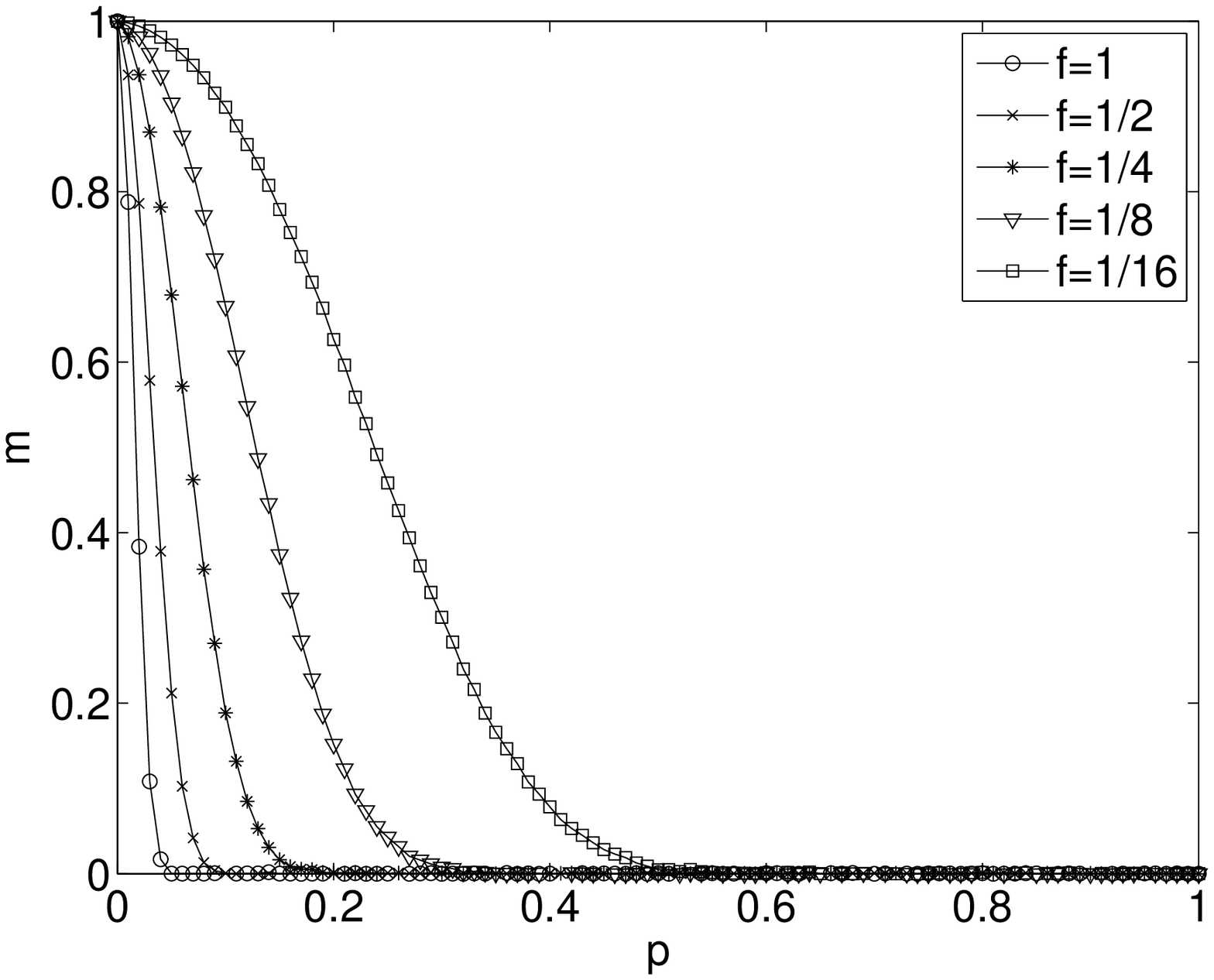}
\includegraphics[scale=0.4]{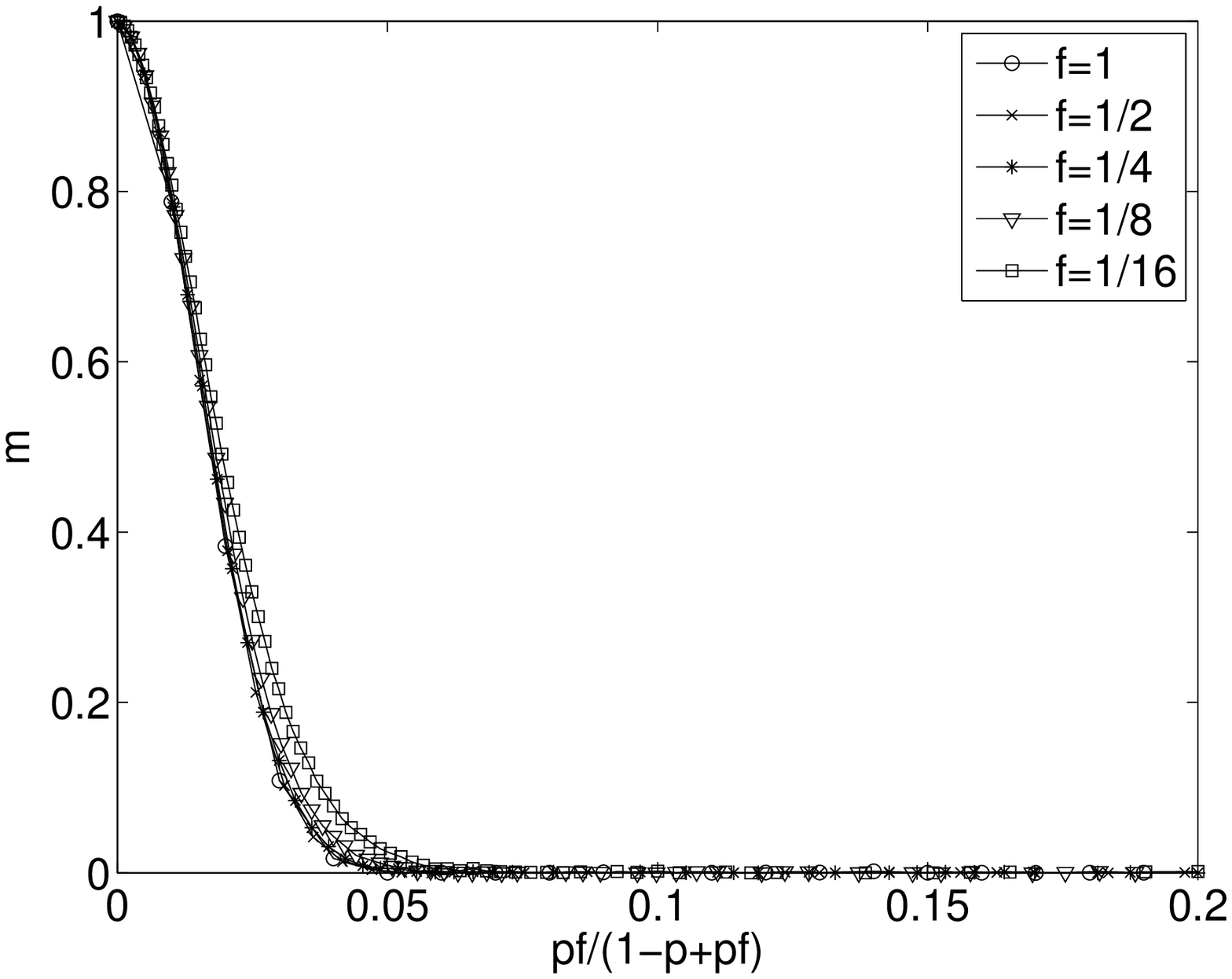}
\caption{Dependence between the magnetization $m$ and the independence factor $p$ for several values of the flexibility factor $f$, for a one dimensional system of size $N=10^4$ (upper panel). The system evolved (`termalized') from an initially ordered state, $m(0)=1$. `Termalization' time was $\tau=500 MCS$ and averaging was done over $10^3$ samples. The scaling found in the case of a complete graph still works (bottom panel).}
\label{D1_scal}
\end{center}
\end{figure}

We consider a chain of length $N$ with periodic boundary conditions. Each site $i=1,\ldots,N$ of the chain is occupied by an Ising spin $S_i=\pm 1$. At each time step we choose randomly a spin $S_i$ and side $s$ ($s=1$ for right, $s=-1$ for left.) The updated state is:
\begin{itemize}
\item conformity (original Sznajd rule), with probability $1-p$: $S_i(t+dt)=S_{i+s}(t)$ if $S_{i+s}(t)=S_{i+2s}(t)$, otherwise $S_i(t+dt)=S_i(t)$,
\item independence, with probability $p$: $S_i(t+dt)=-S_i(t)$, with probability $f$ or $S_i(t+dt)=S_i(t)$, with probability $1-f$.
\end{itemize}
Time $t \rightarrow t+1$ after $N$ elementary time steps, i.e. $dt=1/N$. We have chosen as an initial condition, ferromagnetic order ($m(0)=1$). We made Monte Carlo simulations for several lattice sizes $N=4 \times 10^2,9 \times 10^2 ,64 \times 10^2,10^4,4 \times 10^4$, but here we present results for only one selected lattice size, $N=10^4$. We have measured the magnetization of the system after various `termalization' times $\tau \in [10,10^4]$. The averaging has been done over $10^3$ samples. First of all, we notice that the scaling found in the previous section and given by Eq. (\ref{scal}) is still valid for one dimensional systems -- see Fig. \ref{D1_scal}. 

The results presented in Fig. \ref{D1_scal} suggest the existence of a phase transition, which is unexpected for one-dimensional systems with short-range interactions. However, one should remember that the initial state in our simulations was ordered ($m(0)=1$).
Starting from this ordered state, the system evolves toward a stationary state that depends on the ratio $pf/(1-p+pf)$. After a short `termalization' time, like in Fig. \ref{D1_scal}, the system might be still ordered but finally it will reach its real steady state, which is expected to be disordered. To check the validity of our expectations let us now present the dependence between the magnetization $m$ and the independence factor $p$, for several `termalization' times $\tau$ and a given value of flexibility $f=1/2$. It has been seen that with an increasing $\tau$ the threshold value $p^*$, below which the system is ordered, decreases, suggesting the lack of the phase transition in one dimension. For the infinite system, $N \rightarrow \infty$ and $\tau \rightarrow \infty$, order is present in the system only for $pf/(1-p+pf)=0$. This is an expected result, because only short range interactions are present in the model and the independence $pf/(1-p+pf)$ plays the role of a temperature.
\begin{figure}
\begin{center}
\includegraphics[scale=0.4]{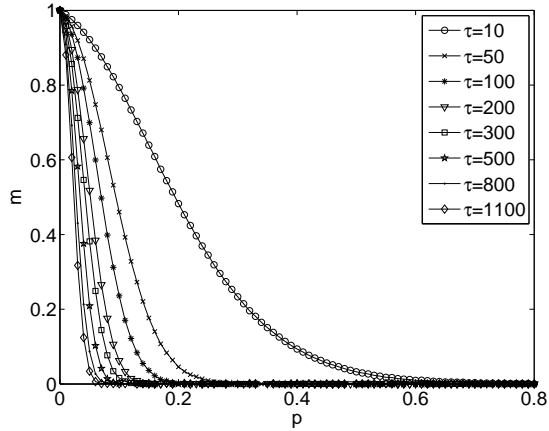}
\caption{Dependence between the magnetization $m$ and the independence factor $p$, for a flexibility factor $f=0.5$, several `termalization' times, $\tau$, and an one dimensional system of size $N=10^4$. The system evolved (`termalized') from an initially ordered state, $m=1$. Averaging was done over $10^3$ samples. With an increasing $\tau$, the critical value $p_c$ is decreasing, suggesting the lack of the phase transition in one dimension.}
\label{D1_term}
\end{center}
\end{figure}

\section{Model on a square lattice}
In this section we consider a square lattice $L \times L$ with periodic boundary conditions. Each lattice site is occupied by an individual, characterized by a binary opinion $S_i=+1$ (in favor) or  $S_i=-1$ (against). In each elementary time step $t$, a $2 \times 2$ box of four neighboring spins is chosen randomly and influences one of the 8 neighboring sites of the box, denoted as $S_i$:

\begin{table}[htbp]
\centering
\begin{tabular}{cccc}
  & $\bullet$ & $\bullet$ & \\
$\bullet$ & $\uparrow$ & $\uparrow$ & $\bullet$ \\		
$\bullet$ & $\uparrow$ & $\uparrow$ & $\bullet$ \\	
  & $\bullet$ & $\bullet$ & 
\end{tabular}
\end{table}

In this paper, we use a modified version of the two-dimensional model introduced to study duopoly markets in \cite{SWW2003}, therefore we update the state in the following way:
\begin{itemize}
\item conformity, with probability $1-p$: 
If all four spins in the box have the same value, they will convince one of the eight nearest neighbors $S_i$, changing its orientation in the direction of the spins in the box. If one of the spins in the box has the opposite orientation to the other three spins, then the neighbor changes its orientation to the orientation of the majority, with probability 3/4. In the case when there is no majority, i.e., two spins in the panel are up and two are down, nothing changes, i.e. $S_i(t+dt)=S_i(t)$.
\item independence, with probability $p$: $S_i(t+dt)=-S_i(t)$, with probability $f$ or $S_i(t+dt)=S_i(t)$, with probability $1-f$.
\end{itemize}

\begin{figure}
\begin{center}
\includegraphics[scale=0.4]{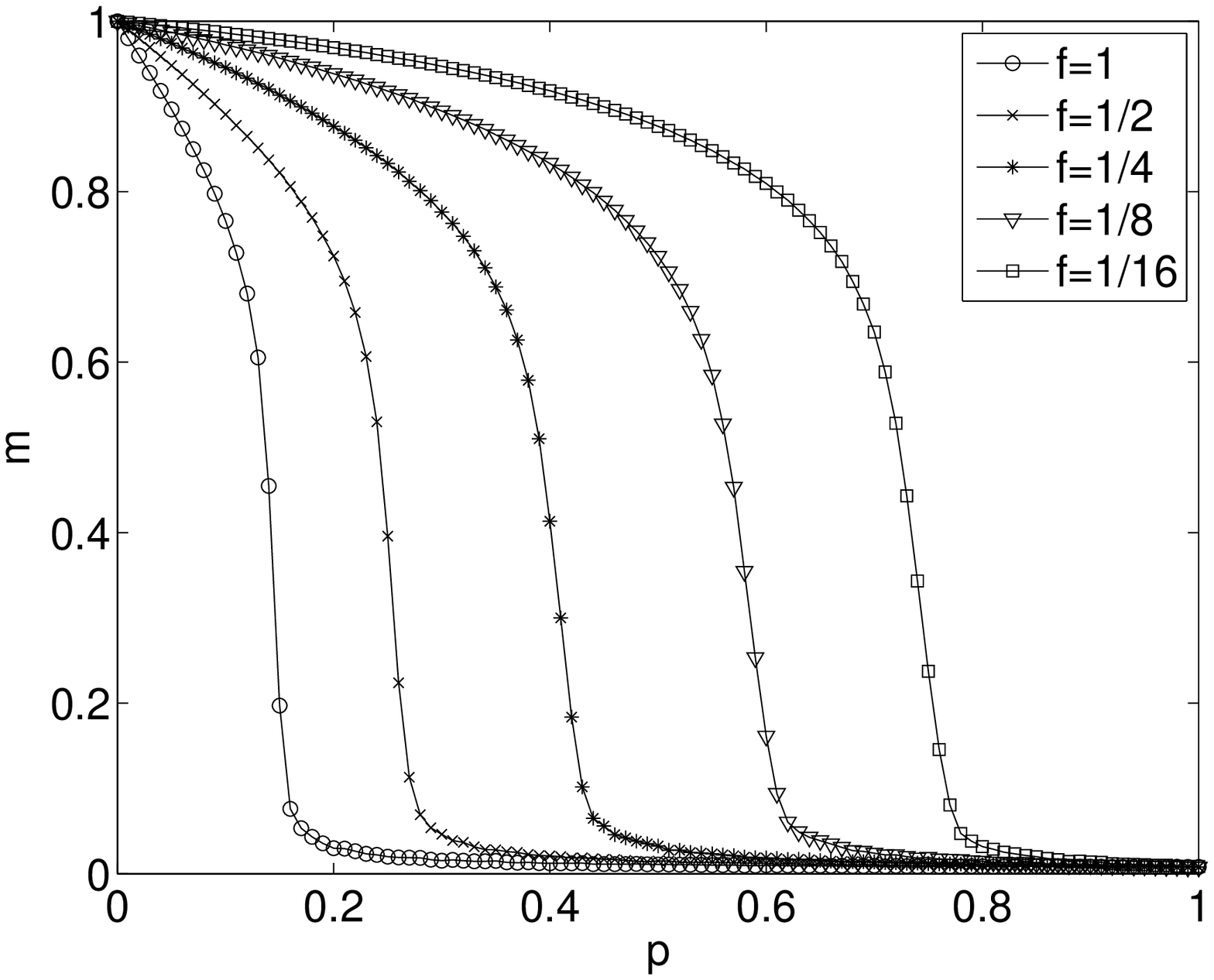}
\includegraphics[scale=0.4]{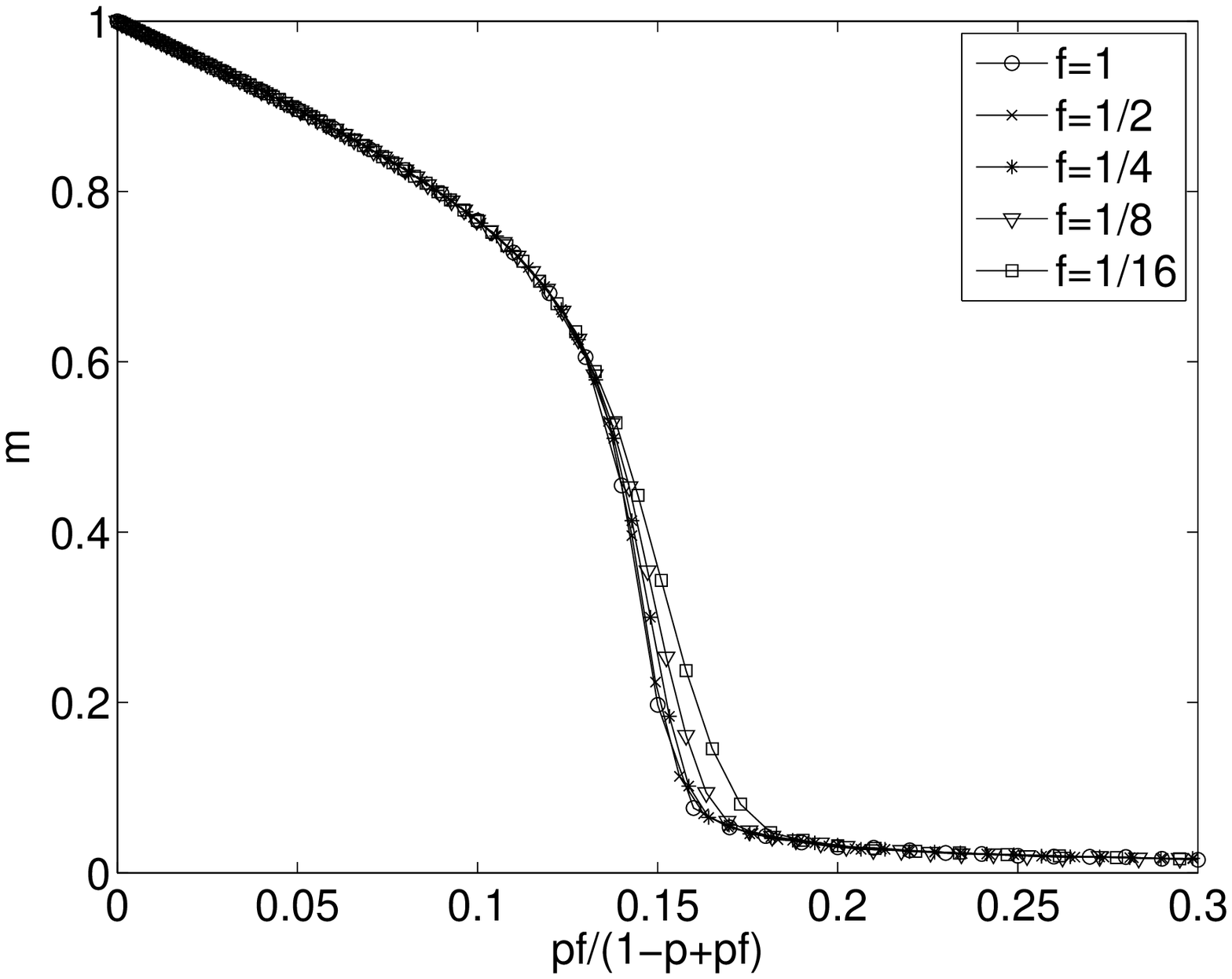}
\caption{Dependence between the magnetization $m$ and the independence factor $p$ for several values of the flexibility factor $f$, in the case of a square lattice, $101 \times 101$. The system evolved (`termalized') from an initially ordered state, $m(0)=1$. `Termalization' time was $\tau=500 MCS$ and averaging was done over $10^3$ samples. The critical value of independence $p_c$ decays with flexibility $f$ (upper panel). The scaling found in the case of a complete graph still works (bottom panel).}
\label{D2_scal}
\end{center}
\end{figure}

\begin{figure}
\begin{center}
\includegraphics[scale=0.4]{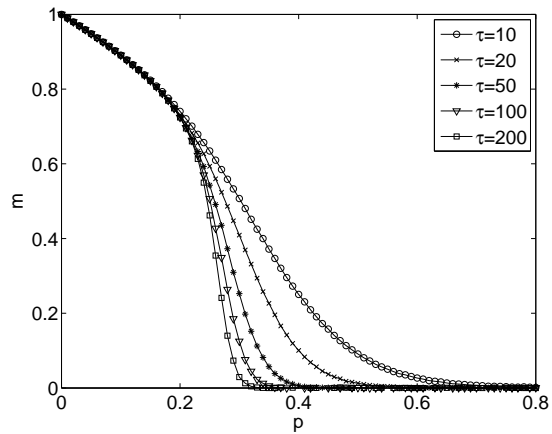}
\caption{Dependence between the magnetization $m$ and the independence factor $p$, for a flexibility factor $f=0.5$, several `termalization' times, $\tau$, and a square lattice, $101 \times 101$. The system evolved (`termalized') from an initially ordered state, $m(0)=1$. Averaging was done over $10^3$ samples.}
\label{D2_term}
\end{center}
\end{figure}

We measure again the magnetization as a function of the independence factor $p$ for several values of the flexibility factor $f$ (see Fig. \ref{D2_scal}). In the case of the square lattice there is no doubt that there is a well defined continuous phase transition (see Fig. \ref{D2_term}) at  $p=p_c$. As in the other cases, scaling (given by eq. (\ref{scal})) is valid and the critical value $p_c$ of independence increases with decreasing flexibility $f$. 

This means that in the case of high $f$ (non-conservative societies) consensus is possible only in the case of low independence (high conformity). For example (see Fig. \ref{D2_scal}), for a relatively conservative society with $f=1/16$ the critical independence is $p_c \approx 0.8$. This means that for a level of independence up to $4/5$ there is a majority in the society. If the society is less `conservative', e.g. $f=1/4$ then the critical value of independence is $p_c \approx 0.5$. This means that if the level of independence were, for example, $p=0.7$ there would be no majority in the system (status-quo), while for the same value of $p$ in the case of $f=1/16$ consensus would be possible. This result could be one of the possible explanations for why the status-quo situation is more and more common in modern societies. Another explanation, connected with the public debate, has been proposed several years ago by Galam \cite{G2004}.

\section{Summary}
In this paper we have introduced a modified version of the original Sznajd model, in which two types of social influence were considered -- independence (with probability $p$) and conformity (with probability $1-p$).  Conformity in our model has been modeled analogously to the Sznajd model, i.e. in the case of conformal behavior, individuals have followed the group norm. Of course, this type of social influence could be modeled also by the voter or majority models. In the case of independent behavior individuals take actions (change opinion) independently on the group norm. Of course, even in the case of independent behavior an individual can change opinion, but it does not depend on the social norm. Therefore, we have introduced a flexibility factor $f$, which denotes the probability of opinion changes in the case of independent behavior. For $f=0$, individuals never change opinion in the case of independent behavior and for $f=1$, the opinion is always changed. From this point of view, varying $f$ we can model the level of conservatism in the society -- it decreases with increasing $f$.

We have studied the model in three cases: complete graphs, one-dimensional systems and two-dimensional square lattices. We found that, in the case of the complete graph and of the two-dimensional system, there is phase transition for a critical value of independence $p_c=\frac{1}{1+\alpha f}$, where $\alpha$ is a constant that depends on the lattice. In the case of a complete graph it can be calculated analytically as $\alpha=4$. Below the critical value of independence, $p<p_c$, the majority coexists with the minority, i.e. the public opinion is $m \ne 0$. Therefore, in the case of a democratic voting, one of the two options wins. For high independence, $p>p_c$, there is a stalemate situation in the society, i.e. $m=0$. In our opinion the most interesting result is the dependence between the critical value of independence and flexibility, $p_c=\frac{1}{1+\alpha f}$ -- the critical value of independence decreasing with increasing flexibility $f$.

This means that in the case of high $f$ (non-conservative societies) consensus is possible only in the case of low independence (high conformity). On the other hand, in conservative societies, even in the case of high independence, consensus is possible. 
In modern societies the value of tradition and hence the level of conservatism seems to be decreasing. This, according to our model,
could be one of the possible explanations for why the status-quo situation observed by Galam \cite{G2004} is more and more encountered now days.



\begin{thebibliography}{10}
\bibitem{CFL2009}
C. Castellano, S. Fortunato and V. Loreto , Rev. Mod. Phys. \textbf{81} (2009) 592
\bibitem{HL1975}
R. Holley and T. Liggett, Ann. Probab. \textbf{3} (1975) 643
\bibitem{G2002}
S. Galam, Eur. Phys. J. B \textbf{25} (2002) 403
\bibitem{KR2003}
P.L.Krapivsky and S. Redner, Phys. Rev. Lett. \textbf{90} (2003) 238701.
\bibitem{SWS2000}
K. Sznajd-Weron, J. Sznajd, Int. J. Mod. Phys. C \textbf{11} (2000) 1157
\bibitem{G2004}
S. Galam S, Physica A \textbf{333} (2004) 453
\bibitem{GJ2007}
S. Galam and F. Jacobs, Physica A \textbf{381} (2007) 366
\bibitem{LLW2005}
M.S. de la Lama, J.M. Lopez and H.S. Wio, Europhys. Lett. \textbf{72} (2005) 851 
\bibitem{Nail_2000}
Nail P, MacDonald G, Levy D, Psychological Bulletin \textbf{126}, 454 (2000)
\bibitem{SBAH2006}
M. R. Solomon, G. Bamossy, S. Askegaard, M. K. Hogg, \emph{Consumer behavior, 3ed}, Prentice Hall (2006)
\bibitem{Myers2010}
D. G. Myers, \emph{Social Psychology, 10ed}, McGraw-Hill (2010)
\bibitem{SL03}
F. Slanina and H. Lavicka, Eur. Phys. J. B {\bf 35}, 279 (2003)
\bibitem{SSP2008}
F. Slanina, K. Sznajd-Weron and P. Przyby³a, EPL \textbf{82} (2008) 18006
\bibitem{SSO2000}
D. Stauffer, A.O. Sousa, M. De Oliveira, Int. J. Mod. Phys. C11, 1239 (2000)
\bibitem{SWW2003}
K. Sznajd-Weron, R. Weron, Physica A 324, 437 (2003) 
\end{thebibliography}
\end{document}